\documentclass[%
 reprint,
superscriptaddress,
 amsmath,amssymb,
 aps,
]{revtex4-1}

\usepackage{graphicx}
\usepackage{dcolumn}
\usepackage{bm}
\usepackage{hyperref}
\usepackage{xcolor}

\usepackage{cases}

\usepackage{comment}

\begin{document}


\title{Flow driven control of pulse width in excitable media}
\author{Adrian Misselwitz}%
\affiliation{Center for Protein Assemblies (CPA) and Department of Biosciences, School of Natural Sciences, Technische Universit\"at M\"unchen, 85748 Garching b. M\"unchen, Germany}
\author{Suzanne Lafon}
 \affiliation{Paris-Saclay University, CNRS, Solid State Physics Laboratory, 91405 Orsay, France}
 \affiliation{Max Planck Institute for Dynamics and Self-Organization, 37077 G\"ottingen, Germany}
\author{Jean-Daniel Julien}%
\affiliation{Max Planck Institute for Dynamics and Self-Organization, 37077 G\"ottingen, Germany}
\author{Karen Alim}
\affiliation{Center for Protein Assemblies (CPA) and Department of Biosciences, School of Natural Sciences, Technische Universit\"at M\"unchen, 85748 Garching b. M\"unchen, Germany}
\affiliation{Max Planck Institute for Dynamics and Self-Organization, 37077 G\"ottingen, Germany}

 \email{k.alim@tum.de}

\date{\today}

\begin{abstract}
Models of pulse formation in nerve conduction have provided manifold insight not only into neuronal dynamics but also the non-linear dynamics of pulse formation in general. Recent observation of neuronal electro-chemical pulses also driving mechanical deformation of the tubular neuronal wall and thereby generating ensuing cytoplasmic flow now question the impact of flow on the electro-chemical dynamics of pulse formation. We, here, theoretically investigate the classical Fitzhugh-Nagumo model now accounting for advective coupling between the pulse propagator typically describing membrane potential and here triggering mechanical deformations and, thus, governing flow magnitude, and the pulse controller, a chemical species advected with the ensuing fluid flow. Employing analytical calculations and numerical simulations we find, that advective coupling allows for a linear control of pulse width while leaving pulse velocity unchanged. We therefore uncover an independent control of pulse width by fluid flow coupling.
\end{abstract}

\maketitle

\section{Introduction}

Neural networks are one of the most widely studied contemporary fields of research. We may untangle the complexities involved in the underlying biology and in the emergent pattern formation due to simplistic yet faithful models for the description of neutral action potentials. The Hodgkin-Huxley model \cite{HodgkinHuxley}, published in 1952, was the first to successfully model action potentials dynamics along the nerve fibre of the squid giant axons. Until today the Hodgkin-Huxley-model is still being used and expanded to accurately describe neural action potentials \cite{borgers, baysal, stiles}. A simplified version of the Hodgkin-Huxley model is the FitzHugh-Nagumo model (FHN), which was developed independently by FitzHugh \cite{FitzHugh} and Nagumo \textit{et. al} \cite{Nagumo} in 1961 and 1962, respectively. The advantage of the FHN model is that it qualitatively retains the non-linear dynamics of the Hodgkin-Huxley model, yet it consists only of two variables, the propagator representing axon membrane potential and the controller describing the chemical species driving propagator dynamics. The reduction to two variables instead of the original four in the Hodgkin-Huxley model, allows for direct analytical insight into the mechanisms of non-linear coupling \cite{hussain, li, mao, Zykov.2018}. Yet, experimental observations challenge the classical description of nerve conduction as the impact of cytoplasmic flows arising from membrane deformation triggered by the pulse are unaccounted for.

The propagation of an action potential along a nerve fiber is accompanied by mechanical deformations of the nerve, including volume expansion and compression \cite{TASAKI_volume, hill_volume}, shortening \cite{TASAKI_tetanic} and a radial change of the nerve fiber \cite{TASAKI_olfactory, TASAKI_olfactory2, TASAKI_swelling, hill_radius}. These mechanical effects are not incorporated in classical Hodgkin-Huxley or FitzHugh-Nagumo models and have, thus, initiated a renewed interest \cite{drukarch_thinking} in model development \cite{Heimburg, RVACHEV_2010}, accounting for example for ensuing fluid flows within the Hodgkin-Huxley model \cite{el_hady} or homogeneous fluid flows within the FHN model \cite{ermakova_propagation_2009, UZUNCA}. Yet, mechanistic insight of how a coupling via fluid flow affects pulse dynamics is missing.

Flow coupling in traveling wave kinetics has, however, been studied in the context of the self-sustained contraction pattern in active porous gels \cite{Radszuweit.2013,Bois.2011,Alonso.2017} and also in tubular geometries \cite{julien}. Mechanistically the flow coupling, here, unfolds as follows \cite{julien}: The gradient of deformation of the tube membrane creates flow. The flow itself creates a flux of chemical species, which in turn affects the gradient of the tube deformation. 

Within the FHN model, we identify the controller as the chemical species concentration and the propagator as the membrane potential. 

We, here, employ analytical derivations and numerical simulations to  investigate the role of an advective coupling arising from mechanical deformations resulting from a propagator and ensuing flows advecting the controller within the FitzHugh-Nagumo-model. We derive a linear dependence of the width of travelling pulses on the advection term and analytically predict the pulse velocity to be independent of the advection term. Both phenomena are corroborated by our numerical simulations, which we further employ to address the effect of model parameters on the impact of the advective coupling. Our results show that advective coupling allows an independent control of pulse width and pulse velocity within the FitzHugh-Nagumo model.

\section{Results}
\subsection{FitzHugh-Nagumo model with propagator-driven advection of the controller}

\begin{figure*}[htbp]
\centering
\includegraphics[width=\textwidth]{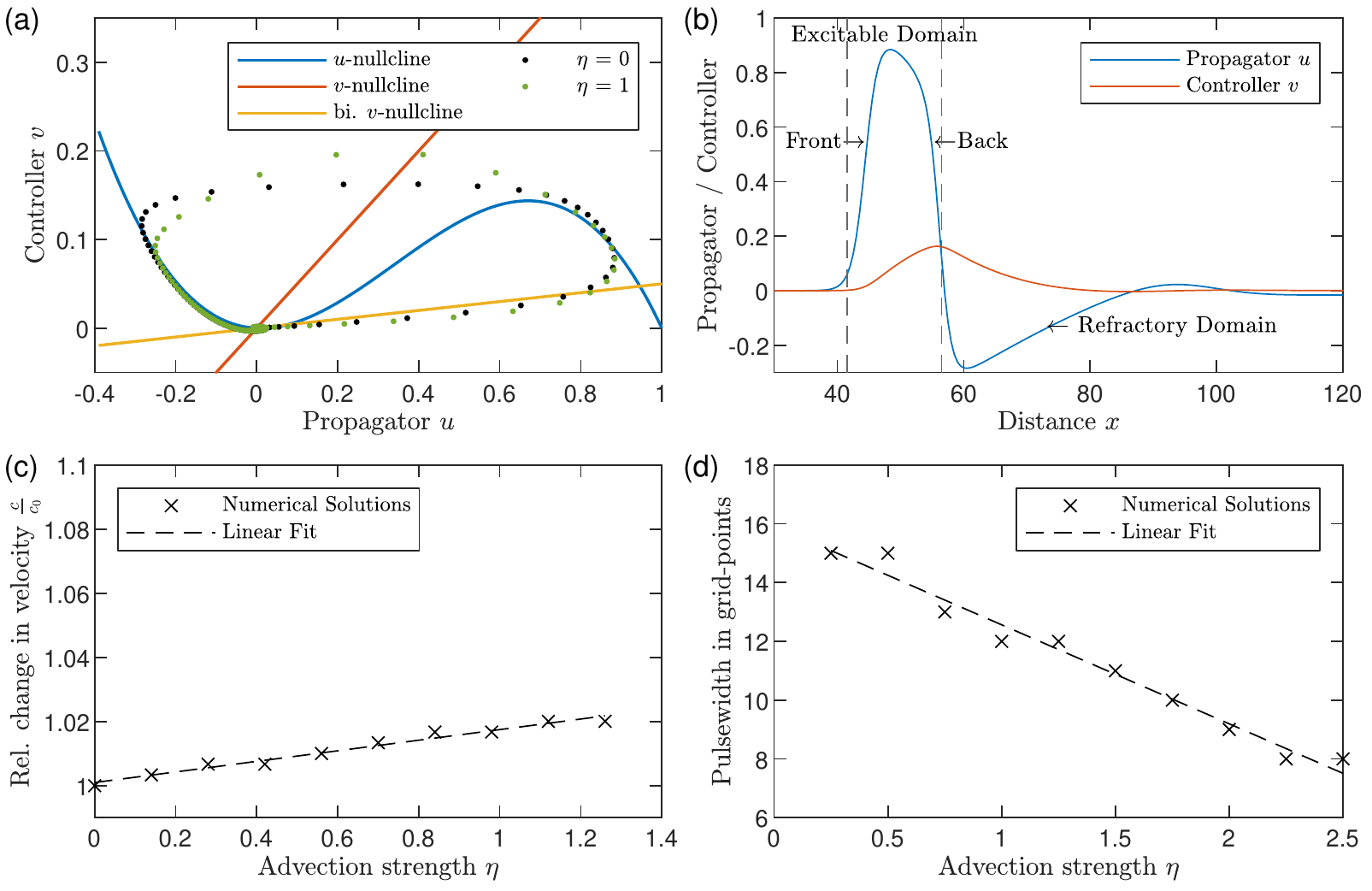}
\caption{Advection coupling leaves pulse velocity unchanged but controls pulse width. a)  Phase portrait of FitzHugh-Nagumo (FHN) without $\eta = 0$ (black, dots) and with advection coupling between propagator and controller $\eta = 1$  (red, dots) along the nullclines (blue - propagator nullcline, red - controller nullcline) being perturbed out of their single stable point. The dots are equitemporal, illustrating the dynamics of the FHN equations. Controller nullcline giving rise to bistability in yellow for reference. 
b) Spatial dynamics of propagator and controller of a leftwards travelling pulse ($\eta=0$) and its subdivision into four phases: front, excited domain, back and refractory domain. Pulse width is defined as the difference between the front,  marked by the initial increase from zero of the controller, and the back, marked by the controller's maximum value (dashed lines).
c) Numerical solutions for the pulse velocity measured as the number of travelled grid-points per unit time. Velocity shows only a small linear dependence (gradient of 0.0082) on advection $\eta$, much smaller than typical variations for changing non-linear parameter $a$.
d) Pulse width decreases linearly with $\eta$. Dashed lines in c) and d) are linear fits of the numerical data. 
System parameters set to $a=0.02$, $b=0.01$, $\gamma = 0.02$ and $D_{u}=0.5$ if not specified otherwise.}
\label{fig:1}
\end{figure*}
To account for the interaction between propagator $u(x,t)$ (membrane potential) and controller $v(x,t)$ (chemical stimulus concentration) arising due to the deformation of the propagator generating fluid flow of magnitude $\eta \frac{\partial}{\partial x}u(x,t)$ which is advecting and dispersing the controller we incorporate an additional advection term that reflects the dependence of the chemical flux on the deformation of the membrane potential $-\frac{\partial}{\partial x}\left(v(x,t)\eta \frac{\partial }{\partial x}u(x,t)\right)$ into the FitzHugh-Nagumo equations,
\begin{subequations}
	\begin{align}
	\frac{\partial u}{\partial t}   &=  D_{u} \frac{\partial^{2}u}{\partial x^{2}} + u(1-u)(u-a)-v
	\label{eq:u-eq_A}\\
	\frac{\partial v}{\partial t}  & = bu - \gamma v - \eta \frac{\partial}{\partial x} \left(v \frac{\partial u}{\partial x} \right)
	\label{eq:v-eq_A}
 	\end{align}
	\label{eq:FitzHughNagumo}
\end{subequations}
where $D_{u} \geq 0$ denotes the diffusivity of the propagator, $\eta$ the viscosity of the fluid, $a > 0$ governs the kinetics of the propagator, and $b \geq 0$ and $\gamma \geq 0$ the kinetics of the controller and thus the stability of the entire system.

Given an initial perturbation of the system, either a travelling pulse or front may form. 
We are, here, interested in the impact of the advection term on the travelling pulse solution. A pulse may form when the the system is monostable, which is the case when the two nullclines of the system defined by $\frac{\partial}{\partial t}u(x,t)==0$ and  $\frac{\partial}{\partial t}v(x,t)==0$ only cross at the point $(0,0)$, as $u=0$ and $v=0$ is the trivial solution of the system of nullclines, see Fig.~\ref{fig:1} a). In the case of three crossings of the nullclines the system is bistable and, thus, may be perturbed out of one of its resting state into a travelling front until it reaches its second stable resting state. Monostability is given when \cite{cross_greenside},

\begin{equation}
	\frac{4b}{\gamma(1-a)^{2}} > 1.
	\label{eq:pulse_cond}
\end{equation}
while bistability is given when the left-hand side of equation \eqref{eq:pulse_cond} is less than $1$. 

 A pulse forms when the system is sufficiently perturbed out of its resting state. Once a pulse is formed, its dynamics are independent of the initial conditions and follow a choreography only dependent on the systems parameters. The dynamics of a pulse can be dissected into four steps, see Fig.~\ref{fig:1} a), b): the pulse \textit{front}, where the system is perturbed out of its stable resting state and the propagator increases quickly until it reaches the vicinity of its nullcline again, the  pulse \textit{excited domain} which corresponds to the peak of the propagator, when dynamics are following the propagators nullcline, the pulse \textit{back} when the propagator drops again sharply back to the other branch of its nullcline and finally the \textit{refractory domain} where dynamics follow again the nullcline and recover back into the stable resting state. This 4-stepped choreography of a travelling pulse remains unchanged under the addition of the advective coupling, see Fig.~\ref{fig:1} a), yet at closer inspection the \emph{dynamics} of the individual steps does seem to be affected by the advection coupling. To gain mechanistic insight on how the advection coupling alters pulse dynamics we turn to analytical derivations on pulse velocity and subsequently pulse length. 
 
\subsection{Pulse velocity derived to be independent of advection strength}

In order to derive closed expressions for the pulse velocity we simplify the dynamical equations Eqs.~\eqref{eq:FitzHughNagumo} by linearly approximating the third order polynomial with a Heaviside function and a linear term, see Ref.~\cite{McKean}. We further incorporate that the dynamics of the propagator $u$ are much faster than the dynamics of the controller $v$ at the \textit{front} and the \textit{back} of the pulse \cite{cross_greenside}, see Fig.~\ref{fig:1} a) by rescaling the fast kinetics of the propagator with the non-dimensional parameter $\epsilon\ll1$ to match time scales of propagator and controller kinetics, together resulting in
\begin{equation}
    \frac{\partial u}{\partial t} = D_{u} \frac{\partial^{2}u}{\partial x^{2}} +\frac{1}{\epsilon} \left[  H(u-a) -u-v\right].
\end{equation}

As we are seeking a travelling pulse solution we only seek solutions where $z=x-ct$ with $c$ being the velocity of the pulse.
In order to remove the explicit $\epsilon$-dependence in the first FHN equation, we introduce the stretching coordinate $\xi =  \frac{z}{\epsilon}$, arriving at
\begin{subequations}
	\begin{align}
	0 = & \tilde{D}_{u} \frac{\partial^{2} u}{\partial \xi^{2}}  + c \frac{\partial u}{\partial \xi}  + H(u-a) -u - v, \\
	0 = & c \frac{\partial v}{\partial \xi}  + \epsilon(bu - \gamma v) - \epsilon \tilde{\eta}\left( \frac{\partial u}{\partial \xi}  \frac{\partial v}{\partial \xi}  + v \frac{\partial^{2} u}{\partial \xi^{2}} \right),
	\end{align}
	\label{eq:FitzHugh-Nagumo_stretch_A}
\end{subequations}
where we rescaled both $\tilde{D}_{u}=D_u/\epsilon^2$ and $\tilde{\eta} = \eta/\epsilon^2$ as both terms describing spatial dynamics should stay unaffected by the unequally fast kinetic terms.
As discussed, at the \textit{front} and \textit{back} of the pulse the dynamics of $\frac{\partial u}{\partial t}$ are much faster than $\frac{\partial v}{\partial t}$, implying very small $\epsilon$. We can therefore evaluate Eqs.~\eqref{eq:FitzHugh-Nagumo_stretch_A} at the \textit{front} and \textit{back} in the limit of $\epsilon \rightarrow 0$. Eq.~\eqref{eq:FitzHugh-Nagumo_stretch_A} (b) then reduces to a  first order differential equation, solved by constant $v$. At the \textit{front} the constant value of the controller equals its stable fixed point value $v=0$, while it takes a finite value of $v=v_{b}$ at the \textit{back} of the pulse, see Fig.~\ref{fig:1} b).
For the \textit{front}, 
\begin{equation}
\tilde{D}_{u} \frac{\partial^{2} u_{f}}{\partial \xi^{2}}  + c \frac{\partial u_{f}}{\partial \xi}  + H \left(u_{f} -a \right)- u_{f} = 0.
\label{eq:front_diff}
\end{equation}
We make the Ansatz $u_{f}(\xi) = C e^{\xi \lambda}$ and require $u_{f}(\xi)$ to converge for $\xi \rightarrow \pm \infty$. Using the jump and continuity condition at $\xi = a$
\begin{subequations}
    \begin{align}
    & A e^{\xi_{a} \frac{-c + \sqrt{c^2+4\tilde{D}_{u}}}{2\tilde{D}_{u}}} = B e^{\xi_{a} \frac{-c - \sqrt{c^2+4\tilde{D}_{u}}}{2\tilde{D}_{u}}} +1, \\
    & A e^{\xi_{a} \frac{-c + \sqrt{c^2+4\tilde{D}_{u}}}{2\tilde{D}_{u}}}= B e^{\xi_{a} \frac{-c - \sqrt{c^2+4\tilde{D}_{u}}}{2\tilde{D}_{u}}} \frac{-c - \sqrt{c^2+4\tilde{D}_{u}}}{-c + \sqrt{c^2+4\tilde{D}_{u}}},
    \end{align}
    \label{eq:jump_cont}
\end{subequations}
we derive the velocity of the pulse at the \textit{front} as
\begin{equation}
    c_{f} = \pm \sqrt{\tilde{D}_{u}} \frac{1-2a}{\sqrt{a(1-a)}}.
    \label{eq:velocity}
\end{equation}
Analogously the dynamics for the propagator at the \textit{back} of the pulse follow from Eq.~\eqref{eq:FitzHugh-Nagumo_stretch_A} to be determined by,
\begin{equation}
\tilde{D}_{u} \frac{\partial^{2} u_{f}}{\partial \xi^{2}}  + c \frac{\partial u_{f}}{\partial \xi}  + H(u_{f}-a)- u_{f} - v_{b} = 0
\label{eq:back_diff}.
\end{equation}
Again employing the Ansatz $u_{b} = C e^{\xi \lambda}$ and respecting jump and continuity condition Eq.~\eqref{eq:jump_cont}
we obtain the velocity for the \textit{back} of the pulse
\begin{equation}
c_{b} = \pm \sqrt{\tilde{D}_{u}} \frac{1-2(a+v_{b})}{\sqrt{(a+v_{b})(1-a-v_{b})}}.
\end{equation}
Note, that we seek solutions where the  shape of the pulse remains constant as it travels through space. This implies that the \textit{front} and the \textit{back} need to travel at the same velocity. Since the \textit{back} is a reversed \textit{front} \cite{cross_greenside}
, we obtain the condition $c_{f} = - c_{b} = c$. This relation of the front and back velocities determines the controller at the \textit{back} of the pulse $v_{b}$
\begin{equation}
v_{b} = 1-2a,
\label{eq:vb}
\end{equation}
and finally the velocity of a travelling pulse
\begin{equation}
	c(\tilde{D}_{u},a) =  \sqrt{\tilde{D}_{u}} \frac{1-2a}{\sqrt{a(1-a)}}.
\label{eq:velocity_pulse}
\end{equation}
Strikingly the advection coupling does not affect the pulse velocity to zeroth order. A result that we indeed confirm in numerical integration of the full set of equations, see Fig.~\ref{fig:1} c).

\subsection{Pulse width analytically predicted to shrink with advection strength}
To analytically derive the pulse width, we aim to solve for the trajectory of the controller $v$. For low order in $\epsilon$ the pulse follows the propagator nullcline during the \textit{excited domain} of the pulse, thereby tracing out the change in controller from $v=0$ at the \textit{front} of the pulse to $v=v_b$ at the \textit{back} of the pulse. Thus, the propagator along the nullcline is given by $u = 1-v$ for $0< v < v_{b}$. The dynamics of the controller along the \textit{excited domain} of the pulse then follow as a function of the spatial coordinate $z$ from Eq.~\eqref{eq:FitzHugh-Nagumo_stretch_A} to be determined by
\begin{equation}
c \frac{\partial v}{\partial z}  - (b+\gamma)v + b + \epsilon^{2}\tilde{\eta}\left(v\frac{\partial^{2} v}{\partial z^{2}}  + \left(\frac{\partial v}{\partial z} \right)^{2} \right) = 0.
\label{eq:diff_A}
\end{equation} 
We can solve the dynamics for the controller at zeroth order in $\epsilon$, i.e.~without the advection term, simplifying the differential equation to
\begin{equation}
c \frac{\partial v}{\partial z}  - (b+\gamma)v + b = 0,
\end{equation}
which is solved by
\begin{equation}
v(z) = \frac{b}{b + \gamma} + Q e^{\frac{b+ \gamma}{c}z},
\end{equation}
where $Q = -\frac{b}{b+\gamma} e^{-\frac{b+\gamma}{c}z_{1}}$ is an integration constant, defined by $v(z_{1}) = 0$. We define the pulse width as the distance travelled during the \textit{excited domain} of the pulse. In the dynamics of the controller this translates to the distance travelled between $v(z_{1}) = 0$ and $v(z_{2}) = v_{b}$ for a rightward travelling pulse. Therefore, the pulse width is given by $\lambda = z_{1}- z_{2}$, following from 
\begin{equation}
v(z_2)=v_{b} = \frac{b -b e^{\frac{b+ \gamma}{c}(z_{2}-z_{1})}}{b+\gamma} = \frac{b -b e^{-\frac{b+ \gamma}{c}\lambda}}{b+\gamma}.
\label{eq_pulseeqn}
\end{equation}
To explicitly solve for the pulse width $\lambda$ we consider the order of magnitude of model parameters. In our simulations, we take $b,\gamma \sim \mathcal{O}(0.01)$, obtaining $c \sim \mathcal{O}(0.1)$ and $\lambda \sim \mathcal{O}(10)$, motivating a
Taylor-expansion in $\frac{b+\gamma}{c}\lambda\ll1$ of Eq.~\eqref{eq_pulseeqn} to first order. Simplifying, we obtain
\begin{equation}
\lambda = \frac{cv_{b}}{b}.
\end{equation}
As we are interested in the effect of the advection term on the pulse width, we now consider Eq.~\eqref{eq:diff_A} to full order in $\epsilon$. Simulations show that $\frac{\partial^{2}}{\partial z^{2}} v(z)$ and $\left(\frac{\partial}{\partial z} v(z)\right)^{2}$ do not change considerably with varying $z$ and we therefore define
\begin{subequations}
	\begin{align}
	C_{1}^{2} \equiv \left(\frac{\partial v(z)}{\partial z} \right)^{2} & \approx \text{constant},\\
	C_{2} \equiv \frac{\partial^{2} v(z)}{\partial z^{2}}  & \approx \text{constant}.
	\end{align}
\end{subequations}
Eq.~\eqref{eq:diff_A} then becomes
\begin{equation}
c \frac{\partial v(z)}{\partial z}  - (b+\gamma)v(z) + b + \epsilon^{2} \tilde{\eta} \left(v(z) C_{2} + C_{1}^{2} \right) = 0.
\end{equation}
The general solution to this first order differential equation is
\begin{equation}
v = \frac{\epsilon^{2} \tilde{\eta} C_{1}^{2} + b}{b+\gamma - \epsilon^{2}\tilde{\eta} C_{2}} + P e^{\frac{b+\gamma - \epsilon^{2}\tilde{\eta} C_{2}}{c}z},
\label{eq:v_sol}
\end{equation}
with $P$ an integration constant, that is defined by $v(z_{1})=0$, obtaining
\begin{equation}
P = - \frac{\epsilon^{2} \tilde{\eta} C_{1}^{2} +b}{b+\gamma - \epsilon^{2}\tilde{\eta} C_{2}} e^{- \frac{b+\gamma - \epsilon^{2}\tilde{\eta} C_{2}}{c}z_{1}}.
\end{equation}
Using $v(z_{2}) = v_{b}$ we obtain
\begin{equation}
v_{b} = \frac{\epsilon^{2} \tilde{\eta} C_{1}^{2} + b}{b+\gamma - \epsilon^{2}\tilde{\eta} C_{2}} \left(1 - e^{ -\frac{b+\gamma - \epsilon^{2}\tilde{\eta} C_{2} }{c}\lambda} \right).
\label{eq:vb}
\end{equation}
We can rewrite Eq.~\eqref{eq:vb} as
\begin{equation}
e^{\frac{\lambda}{c} \left(b+\gamma - \epsilon^{2}\tilde{\eta} C_{2} \right)} = \frac{1}{1-\alpha},
\label{eq:alpha}
\end{equation}
with $\alpha = v_{b}\frac{b+ \gamma-\epsilon^{2}\tilde{\eta} C_{2}}{b+\epsilon^{2}\tilde{\eta} C_{1}^{2}}$.
Because the term $\epsilon^{2} \tilde{\eta} C_{2}$ is very small and generally $v_{b} < 1$, see Fig.~\ref{fig:1} b), we obtain\\
\begin{equation}
    \alpha \approx v_{b} \frac{b+\gamma}{b} < 1.
\end{equation} 
Taking the logarithm of Eq.~\eqref{eq:alpha} and assuming $\alpha$ to be sufficiently small
\begin{equation}
\frac{b+\gamma - \epsilon^{2} \tilde{\eta} C_{2}}{c}\lambda = \text{ln}(1) - \text{ln}(1-\alpha) \approx \alpha.
\end{equation}
Solving for $\lambda$ we obtain
\begin{equation}
\lambda = \frac{c v_{b}}{b + \epsilon^{2}\tilde{\eta} C_{1}^{2}} \approx \frac{cv_{b}}{b} \left( 1 - \frac{\epsilon^{2}\tilde{\eta} C_{1}^{2}}{b} \right).
\end{equation}

Assuming that $ C_{1} \approx \frac{\partial}{  \partial z} v\left(z_{1}\right)$, Eq.~\eqref{eq:v_sol} yields
\begin{equation}
C_{1} \approx \frac{\partial}{\partial z}  v\left(z_{1}\right)=  \frac{b+\gamma - \epsilon^{2}\tilde{\eta} C_{2}}{c} \frac{\epsilon^{2} \tilde{\eta} C_{1}^{2} +b}{b+\gamma - \epsilon^{2}\tilde{\eta} C_{2}}.
\end{equation}
Simplifying we obtain
\begin{equation}
C_{1} \approx \frac{\epsilon^{2} \tilde{\eta} C_{1}^{2} +b}{c},
\end{equation}
which is solved by
\begin{equation}
C_{1} = \frac{-c \pm \sqrt{c^{2} - 4b\epsilon^{2}\tilde{\eta}}}{2\epsilon^{2}\tilde{\eta}}.
\end{equation}
Because $\epsilon^{2} \ll 1$, we Taylor-expand $C_{1}$ to first order, 
\begin{equation}
	C_{1} \approx -\frac{b}{c},
\end{equation}
 and, thus, obtain a closed expression for the width of a pulse
\begin{equation}
\lambda = \frac{c(1-2a)}{b}\left( 1 - \epsilon^{2} \frac{b\tilde{\eta}}{c^{2}} \right).
\label{eq:width}
\end{equation}
Thus, we find that in contrast to the pulse velocity, the pulse width is affected by advective coupling, see Fig. \ref{fig:1} d). The higher the fluids viscosity $\tilde{\eta}$ the smaller the pulse width. To test the validity of the analytical expressions we next turn to numerical integration of the full set of Eqs.~\eqref{eq:FitzHughNagumo}.

\subsection{Implicit integration of advection coupled dynamics system required for stability}
To integrate excitable media dynamics with an advection term we employ a $\theta$-weighted Crank Nicolson scheme \cite{Hoeve_2010, Shi_1994, eggers_dupont_1994}. The algorithms' basic structure follows that of a Newton method, but differs from it by dynamically adjusting the time steps and evaluating the dynamic equations not at a time step $i$ but at time steps $i+\theta$. To illustrate the basics of the algorithm, we consider the general set of differential equations
\begin{equation}
\frac{\partial}{\partial t} \vec{y} = \vec{f} \left(\vec{y} \right),
\end{equation} 
which correspond to Eqs.~\eqref{eq:FitzHughNagumo} in our implementation. We denote $y^{n}_{i}$ to be the variable $y$ at time $n$ and on grid point $i$, with $n \in \{0,t_{f} \}$, for some final time $t_{f}$ and $i \in \{ 0, N\}$, for some number of equally spaced grid points $N$ in the one-dimensional system.
We now consider the residual that we obtain, when approximating the function $\vec{f}(\vec{y})$ to linear order:
\begin{equation}
\vec{r} \left(\vec{y}^{n+1} \right) = \frac{\vec{y}^{n+1} - \vec{y}^{n}}{\Delta t} - \vec{f} \left(\vec{y}^{n+\theta} \right),
\label{eq:residual}
\end{equation}
with $0\leq \theta \leq 1$.
For $\theta=0$ we obtain a fully explicit and for $\theta =1$ a fully implicit method. For our simulations we will take $\theta = 0.55$, as it has been found to improve stability at the cost of only slightly less accuracy \cite{eggers_dupont_1994}.
We are looking for values of $\vec{y}^{n+1}$ such that the residuals become $\vec{r} =0$. For this we use Newton's method. A good initial guess is assuming $\vec{y}^{n+1} - \vec{y}^{n}$
to be equal to $\vec{y}^{n} - \vec{y}^{n-1}$. In order to dynamically adjust the efficiency of the simulation, we allow time-steps to vary in magnitude, obtaining:
\begin{equation}
\vec{y}^{n+1}_{est} \approx \vec{y}^{n} + \left(\vec{y}^{n} - \vec{y}^{n-1} \right) \frac{\Delta t_{n\rightarrow n+1}}{\Delta t_{n-1 \rightarrow n}}.
\label{eq:du}
\end{equation}
We correct our initial estimate by subtracting the inverse of the product of the Jacobian, a matrix containing all first order derivatives of every grid point, and the residuals of our estimate,
\begin{equation}
\vec{y}^{n+1} = \vec{y}^{n+1}_{est} - \textbf{J}^{-1} \vec{r} \left(\vec{y}^{n+1} \right).
\label{eq:Jacobian}
\end{equation}
The inversion of the Jacobian is the costliest part of the algorithm, as we use a grid with order $\mathcal {O}(1000)$ grid points. Because of this, we only make one correction with the Jacobian per iteration. To ensure, that this correction is sufficient, we ensure, that our initial guess is not far off from the actual value, by keeping the time steps $\Delta t_{n\rightarrow n+1}$
small. This adjusting can be done automatically, by letting the algorithm calculate the time step thrice. Once for a step size $\Delta t$ and twice successively for step size $\frac{\Delta t}{2}$. The two half steps will result in a more accurate approximation. If the relative error between the two steps is smaller than a threshold $\chi$ for say 10 steps, we increase the step size by a factor $2^{\frac{1}{4}}$, reducing the calculation cost without sacrificing much accuracy. If the error is larger than threshold $\chi$ we decrease the step size by a factor $\frac{1}{2}$, ensuring a good accuracy of the simulation.

\subsection{Limitations of pulse generation in parameter space due to advective coupling} 
\begin{figure*}[ht]
\centering
    \includegraphics[width=\textwidth]{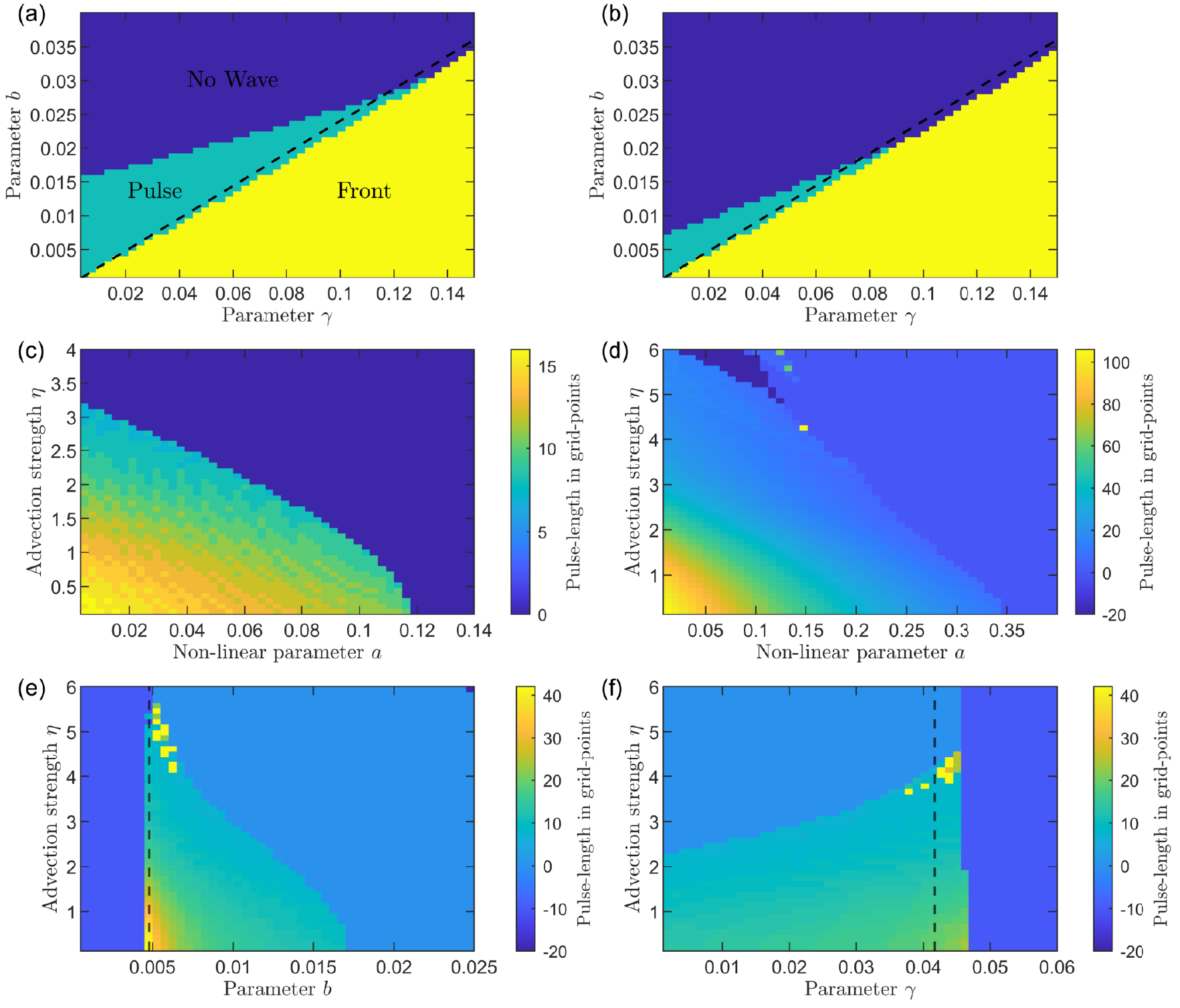}
\caption{Pulse formation and pulse width as function of system parameters. (a), (b) Pulses (green) form above analytical predicted condition for monostability (dashed lines, see Eq.~\eqref{eq:pulse_cond}), separating them from travelling fronts. The upper limit for pulse formation shrinks as advection coupling compare zero  $\eta=0$ in (a) and finite coupling $\eta=3$ in (b), respectively. Blue indicates parameters where no travelling pulse forms. (c) and (d) Pulse width as a function of  $\eta$ and $a$. The blue region indicates no pulse generation. (c) Shows the space for $b=0.01$ and $\gamma=0.02$ and (d) for $b=0.001$ and $\gamma=0.002$. (e) Pulse width as function of parameters $\eta$ and $b$, with $a=0.02$, $\gamma = 0.02$. (f) Pulse width as function of parameters $\tilde{\eta}$ and $\gamma$, with $a=0.02$, $b = 0.01$.
In both (e) and (f) the yellow points at large $\eta$ correspond to numerical artifacts. The vertical cut-off at small $b$ and large $\gamma$, respectively, indicates the transition to front (dashed line, see Eq.~\eqref{eq:pulse_cond}). The front forming phase space is indicated by wavelengths of $\lambda = -20$.}
\label{fig:Parameterspace}
\end{figure*}
Before employing our numerical scheme to assess the impact of advective coupling on pulse dynamics we first sweep the parameter space numerically to identify when pulses form. There is a clear cut-off of pulse formation at the transition from monostability to bistability and thus front formation given by Eq.~\eqref{eq:pulse_cond}. Further, from the calculation of pulse velocity Eq.~\eqref{eq:velocity_pulse}, we find that the pulse velocity vanishes as $a$ approaches $\frac{1}{2}$, additionally establishing an upper  theoretical limit $a < \frac{1}{2}$. \\
For all simulations we have taken the diffusion constant to be $D_{u} = 0.5$. Sweeping parameter space $b-\gamma$ at fixed $a=0.02$ we first of all recover the analytic prediction of the transition between front and pulse, see Fig.~\ref{fig:Parameterspace} a). Increasing fluid viscosity $\eta$ and therefore advective coupling keeps the transition to front formation unaffected yet reduces the parameter space for travelling pulses, see Fig.~\ref{fig:Parameterspace} b). Inspection of pulse trajectories in phase space, see Fig.~\ref{fig:1} a), suggest that advective coupling decreases pulse formation as the advective term positively reinforces the controller, which in turn reduces the propagator. The reduced propagator switches earlier from the \emph{excited domain} to the \emph{back}, resulting in a narrower pulse. \\
To explore the limit on parameter $a$ we sweep the $\eta$-$a$ parameter space for $b = 0.01$, $\gamma = 0.02$, see Fig.~\ref{fig:Parameterspace} c). We observe pulse generation only for small values of $\eta$ and $a$ and note a cut-off at about $a=0.12$, a value much smaller than the theoretical limit. Yet, decreasing the magnitude of $b$ and $\gamma$ by one order, significantly increase the parameter space for pulse generation, see Fig.~\ref{fig:Parameterspace} d). Even smaller values of $b$ and $\gamma$ achieve pulse generation for $a$ close to $0.5$. The pulse generation in this regime is limited by numerical instabilities. As a one order magnitude decrease in $b$ and $\gamma$ increases the pulse width by roughly one order of magnitude, the relative changes of the propagator and controller between grid points becomes smaller, ensuring stability over a larger parameter space. \\
\begin{figure*}[ht]
\centering
    \includegraphics[width=\textwidth]{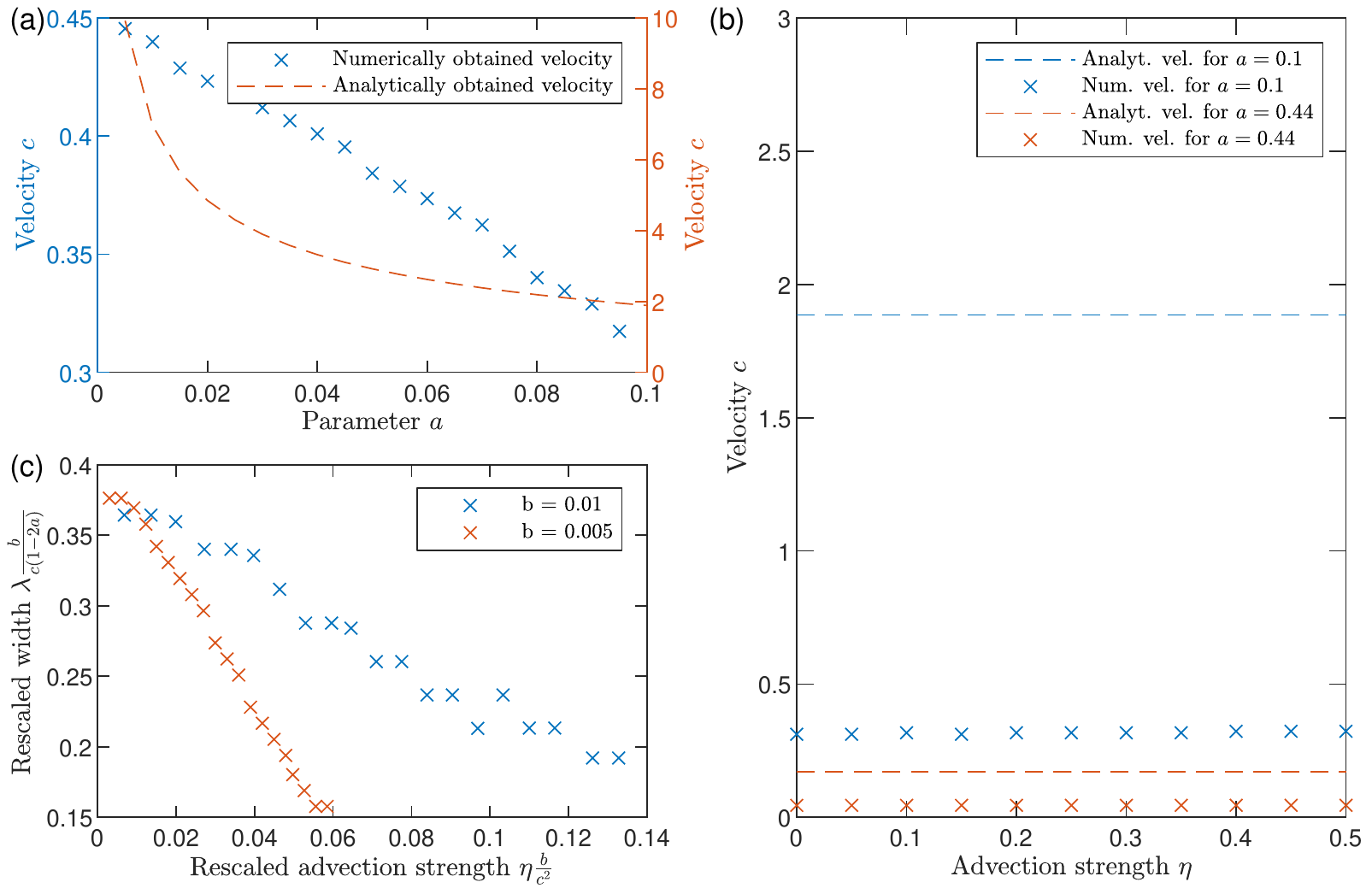}
\caption{Comparison of analytical and numerical pulse velocity and pulse width reveals success of analytical prediction of pulse width governed by advection strength. (a) The numerical and analytical, see Eq.~\eqref{eq:velocity_pulse}, pulse velocity as function of parameter $a$. Both coincide well for large values of $a$, as both vanishing for $a\rightarrow 0.5$, but the analytical result predicts a divergence for small $a$. (b) The pulse velocity shows a small but neglectable dependence on the advection strength $\eta$ as predicted analytically, which is maintained for small and large values of $a$. (c) Pulse width dependence on $\eta$ with rescaling factors from Eq.~\eqref{eq:width} versus numerical results. As predicted, there is a linear dependence between $\lambda$ and $\eta$, however the slope is not unitary, but rather strongly dependent on other model parameters like $b$.}
\label{fig:vel_width}
\end{figure*}
Even though a bit more convoluted, the above arguments also explain pulse formation along the sweeps through the parameter space spanned by $\eta$-$b$ and $\eta$-$\gamma$, see Figs.~\ref{fig:Parameterspace} e), f) respectively. Here the clear cut-off at the transition from pulse to front formation is again exemplified. We here numerically explored the pulse formation broadly within the parameter space and next turn to explicitly test our analytical prediction on pulse velocity and pulse width as a function of advection strength.

\subsection{Pulse width governed by advective coupling while pulse velocity unaffected}
According to our analytical results advection strength has disparate impacts on pulse velocity and pulse width. Pulse velocity is predicted to be independent of advection strength, see Eq.~\eqref{eq:velocity_pulse}, while pulse width is derived to linearly decrease with advection strength, see Eq.~\eqref{eq:width}. Indeed our numerical results confirm that pulse velocity is well-described as being independent  of advection strength  see Fig. \ref{fig:vel_width} b). We however find that the analytical prediction overestimates the precise value of the pulse velocity. Mapping out in particular the analytically obtained pulse velocity as a function of model parameter $a$, see Fig.~\ref{fig:vel_width} a), we find, that the analytical and numerically obtained velocities agree in the limit of $a\rightarrow0.5$, where both decay to zero, yet a divergence for vanishing $a$ is predicted analytically. Therefore the analytical pulse velocity agrees best with numerical simulations for large $a$. This is to be consistent as we approximated the third order polynomial $f(u,v)$ with a Heaviside function $H(u -a)$. The approximation works best for $a$ close to $0.5$, explaining disagreements in the pulse velocity between analytical and numerical results for small $a$. \\
Assessing the pulse width functional dependence on advection strength we numerically confirm that the pulse width scales linearly with $\eta$, see Fig.~\ref{fig:vel_width} c). This holds for a varying system parameters. We note that the numerical simulations show the system parameters to have a strong effect on the gradient and the $y$-intercept in Fig.~\ref{fig:vel_width} c), which is not captured by the prefactors of the analytical result in Eq.~\eqref{eq:width}. We have found no discernible trend for these effects, yet for all observed parameter ranges, the linear dependence on $\eta$ remained. We note a trend of decreasing pulse width for increasing $b$ and decreasing $\gamma$, in accordance with their positive or negative impact on controller dynamics, see Eq.~\eqref{eq:FitzHughNagumo}. The deviations between analytical and numerical results regarding the model parameters $b$ and $\gamma$ are therefore likely to stem from $v=0$ at the pulse \textit{front} not being fully fulfilled numerically. That said, the functional prediction on the impact of advection strength is unaffected by these quantitative differences. 
\subsection{Pulse generation for negative coupling of the advection term}
\begin{figure}[htbp]
		\centering
		\includegraphics[]{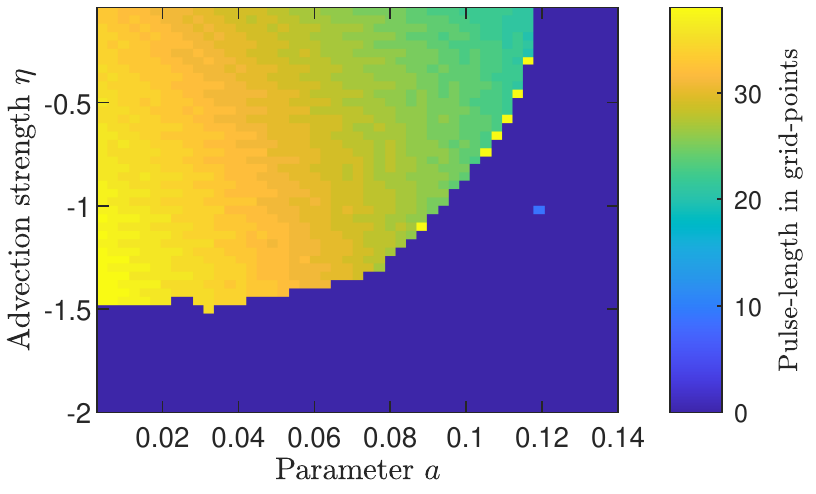}
		\caption{Pulse width for negative advective coupling in $\eta$-$a$ parameter space. The dark blue area indicates parameter-sets resulting in no pulse generation representing either "no travelling wave" or numerically instability. The yellow points indicate numerical artefacts, arising due to the formation of wave trains and other unexpected wave formations at the edge of the pulse generating area.}
		\label{Fig:negative_A}
\end{figure} 
To give a holistic insight into our model, we also want to discuss the effect of negative values of $\eta$. This parameter range does not hold for the bio-mechanical motivation of the advection term, however as the focus of this paper is to present a model, that allows for dynamical changes to a usually constant pulse width, it is worth considering its effect on the whole theoretical parameter range. \\
Following a naive consideration of the linear dependence of pulse width on advection strength, we would expect to find a linear increase of the pulse width for larger negative values of $\eta$. Numerical simulations, shown in Fig.~\ref{Fig:negative_A}, indicate that this is indeed the case, however with a change of pulse width much smaller than for positive $\eta$. The difference in gradients can be explained by the effect that the advection term has on the controller. For positive values, it reinforces the controller by steepening its peak, which in turn increases the magnitude of the advection term, due to its dependence on $\frac{\partial v}{\partial x}$. Negative coupling values result in a reduced controller, leading to a split into two peaks for large enough values of $\eta$. The decreased gradient of the controller results in a weaker advection term, explaining the smaller gradient for negative coupling. \\

In Fig.~\ref{Fig:negative_A} we further see that the parameter-space for pulse solutions is confined to small values of $\eta$ and $a$. For larger $a$ we observe a transition into the state of "no travelling wave", while for larger negative $\eta$ we observe numerical instabilities. While methods such as decreasing the initial amplitude and having a system with an even number of grid-points help numerical stability, we still observe a numerical limitation of the parameter space. The theoretical upper limit for the a pulse width is the size of the system, resulting in an upper limit of $\eta$ that scales with the system size, however our numerical simulations are unable to remain stable for large pulse widths rendering this limit beyond the scope of the present work. \\
\section{Conclusion}
In this paper, we have shown that accounting for advection coupling in the FitzHugh-Nagumo equations leads to novel qualitative properties of its travelling pulse solutions. While the velocity of a pulse is independent of the advection term, the pulse width is now tunable, changing linearly with the coupling parameter of the advection term. 

Our simulations have shown that one can reliably generate pulses for a large area of the parameter space, however with a different order of magnitude of the gradient than predicted analytically. Lastly, we numerically demonstrated the linear dependence of the pulse width for negative coupling.\\
Our model allows for a wider application of the standard FitzHugh-Nagumo model now incorporating flow-based advection of the controller species thereby accounting for mechanical changes of nerve fibers driving fluid flows under action potentials. The additional degree of freedom to adjust the pulse width by modulating the advection strength in these systems may help to form more comprehensive models.

\section*{Acknowledgements}
This work was supported by the Max Planck Society, the Deutsche Forschungsgemeinschaft via FOR-2581 (P1) and the Human Frontiers Science Program via RGP0001/2021-203.


\end{document}